\documentclass[usenatbib]{article}
\usepackage{amsmath}
\usepackage{amssymb}
\usepackage{graphicx}
\usepackage{epsfig}
\usepackage{cite}
\usepackage{natbib}
\usepackage{textcomp}
\title{Reflections on works by I.S. Shklovsky regarding the nature of 
radio galaxies\footnote{\tt Submitted in Astronomical \& Astrophysical Transactions}}
\author{B.V. Komberg\footnote{bkomberg@asc.rssi.ru},\; V.I. Zhuravlev\footnote{zhur@asc.rssi.ru}\\[15pt] 
Astro Space Center of Lebedev Physical Institute, \\  Profsoyuznaya 84/32, Moscow 117997, Russia}
\begin{document}
\maketitle

\begin{abstract}
The paper is a brief overview of the works by Iosif S. Shklovsky (1916--1985), carried out over almost 30 years 
(1955--1985), on the nature of activity (primarily in the radio frequency range) in nuclei of some galaxies.

Worthy of note is Shklovsky's pioneering work of 1962, in which he made an attempt to consider possible 
evolutionary tracks of extragalactic radio sources by constructing an analog of the Herzsprung--Russel 
diagram for stars (radio luminosity at 160 MHz was taken instead of optical luminosity; total radio size 
at the same frequency, as the other parameter). Later works by other authors are also discussed, where 
similar diagrams were plotted using a larger observational material. 

Special attention is paid to the evolution of Shklovsky's views regarding the possible ways of gas getting 
into radio galaxies' central regions, followed by   high-velocity ejections of magnetized plasmons  from 
their nuclei. In his assumptions, Shklovsky was mainly based on the observational data for the properties 
of the  closest radio galaxy, NGC 4486 (Virgo A, M87), which he believed to be the same reference standard 
for extragalactic radio astronomy as the Crab Nebula for galactic radio astronomy. 

Shklovsky's approach to the recurrence of the activity phenomenon in galactic nuclei and the one-sided 
character of radio ejections from them is discussed. 

Modern views on these issues are also briefly considered.    

\end{abstract}


\section{Introduction}
Starting from the mid-1950s, Iosif S. Shklovsky (ISS) repeatedly addressed the issue of the nature of radio emission 
of some galaxies, which in the case of $L_{\rm r}>L_{\rm opt}$ were termed radio galaxies (RG). Here is an 
incomplete list of his works on the subject over 10 years (1955--1964):
\begin{itemize}
\item On the nature of the radio emission from NGC 4486  \citep{sh6},
\item On the nature of a radio source in Sgr A  \citep{sh7},
\item Radio galaxies 
\citep{sh8},
\item The problem of radio galaxies \citep{sh9},
\item On the nature of radio galaxies  \citep{sh10},
\item On the nature of ejections from radio galaxies  \citep{sh11},
\item The new about radio galaxies  \citep{sh12}.
\end{itemize}

In some of his works, Shklovsky criticizes a number of hypotheses put forward on this subject 
at that time. 
For instance, in \cite{sh1}, with respect to the work ``On the identification of radio sources'' by 
\cite{baa}
or in \cite{min}  relative to the work by
Hoyle. In these works, one can see his departure from the assumption that strong radio emission in 
galaxies can be related to processes driven by their collisions. The hypothesis on the relation 
of the phenomenon of strong radio emission from galaxies to the high rate of supernova explosions in 
their central regions does not, in his mind, stand up to criticism, either. Even with account for the 
assumption of the possible ``chain reaction'' of such outbursts \citep{ber}. 

ISS dwells on these issues in his large work \citep{sh9} and also in \cite{sh10}. 
We will discuss these works in greater detail.  

\section {The dependence of $L_{\nu} \sim  R_{\nu}$ for strong radio sources}
\begin{figure}       
\includegraphics[width=0.9\textwidth,angle=0]{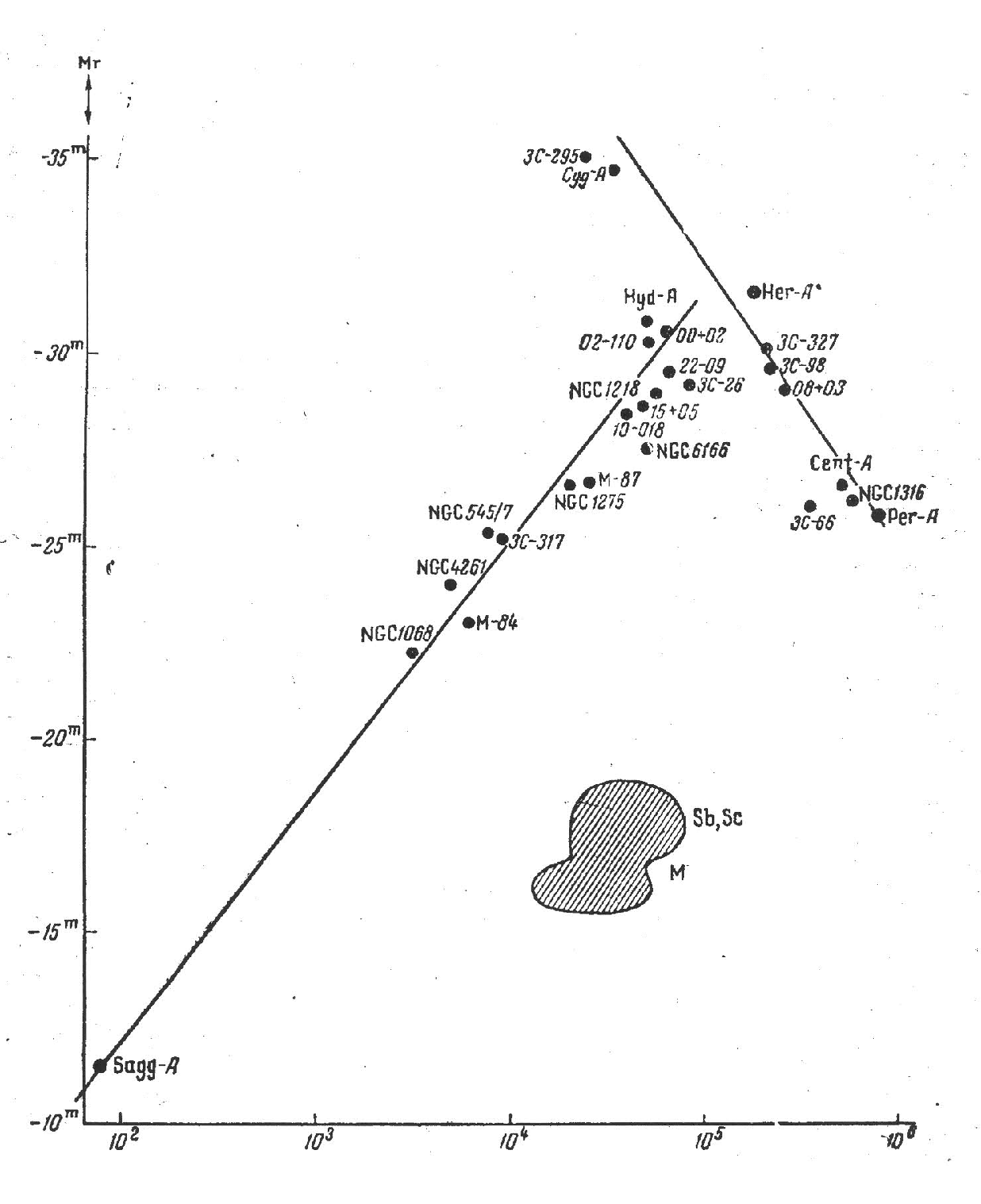}
\caption{Absolute radio magnitude $M_{\rm r}$ vs linear size $l_{\rm r}$ (in pc): 
the increasing dependence, the main sequence; the decreasing dependence, 
the  giant sequence \citep{sh10}. }
\label{f1}
\end{figure}

In these works, Shklovsky attempts to find  a classification scheme for the then known strong radio sources 
(RS) by constructing for them an analog of the Herzsprung--Russel diagram for stars. As 
the most suitable coordinates, he chose the absolute radio magnitude $M_{\rm r} = m_r+5-5\log r$, 
where  $m_r=-53.4-2.5\log F_{160}$, and the  size ($R_{\rm r}$) of radio sources at a frequency of 160 MHz 
(from the literature) (Fig.~\ref{f1}).

\begin{figure}    
\includegraphics[width=0.9\textwidth,angle=0]{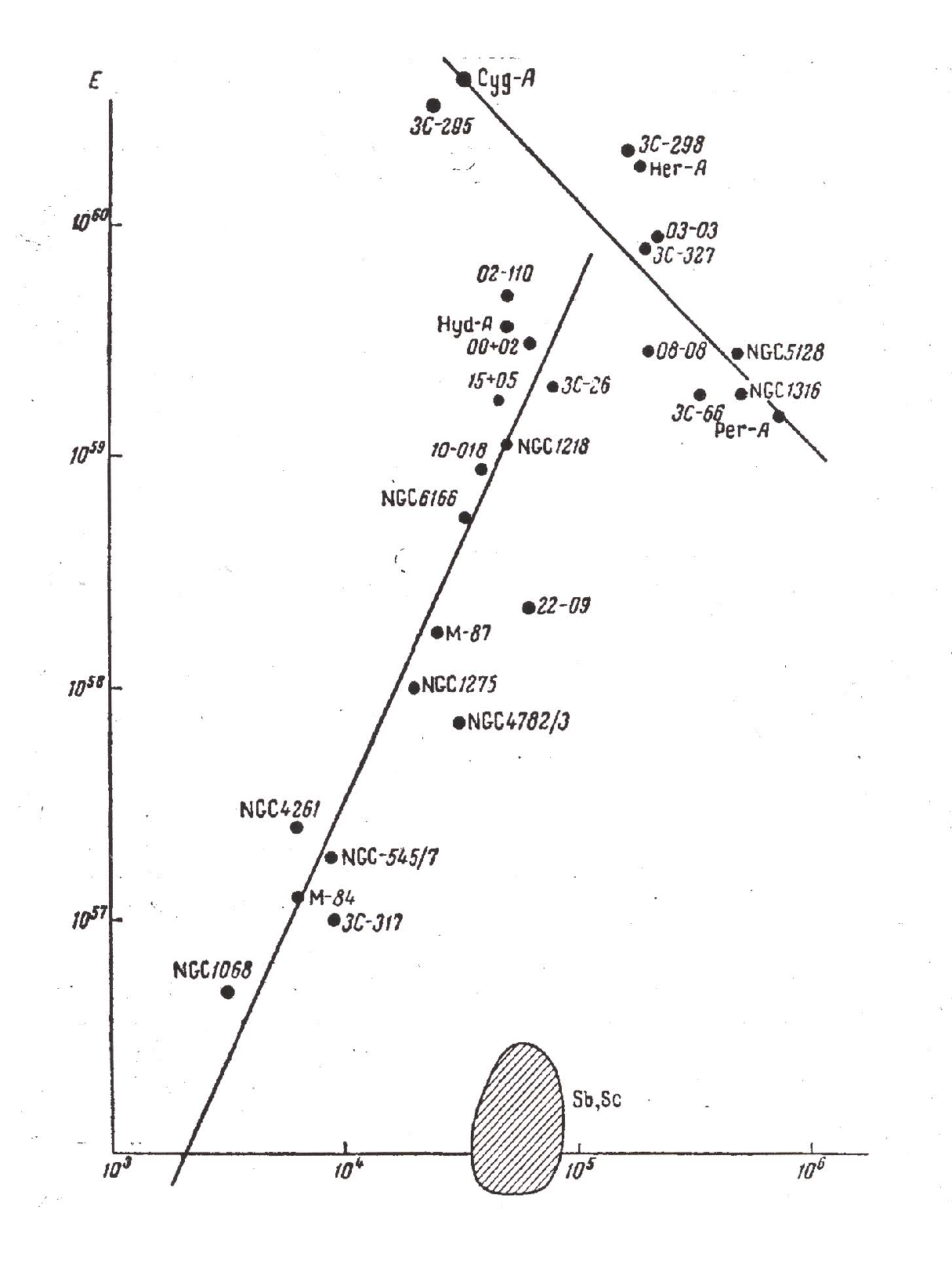}
\caption{The main sequence (increasing curve) and the  
giant sequence (decreasing curve) on the plane: the total energy of relativistic 
particles and magnetic fields ($(E_{\nu}) - l_{\rm r}$ (pc)  \citep{sh10}. }
\label{f2}
\end{figure}

By plotting the data of 26 extragalactic RS and a radio source at the centre of our Galaxy (
Sgr A
) 
using
these coordinates, 
two different sequences are singled out in the opinion of the author: the ``main sequence'' 
on which radio luminosity 
$L_{\nu} \sim R_{\nu}^{2.5}$ (where the mean radiation of a unit volume slowly decreases with the 
volume increasing) 
and the ``giant sequence'' on which $L_{\nu} \sim R_{\nu}^{-4.8}$. The same 
pattern of two sequences 
of RS is found if, instead of the value of $L_{\nu}$, that of the total energy ($E$) of relativistic 
particles and  magnetic field concentrated in extended radio components is plotted: 
$E_{\nu}\sim F_{\nu}^{4/7} r^{17/7}\varphi^{9/7}$, where $r$ is the distance to an RS and 
$\varphi$ are its total angular sizes (Fig.~\ref{f2}). In this case, the main sequence
 is characterized by the dependence $E_{\nu}\sim R_{\nu}^2$, and the giant sequence, by $E_{\nu}\sim R_{\nu}^{-1}$.

In Shklovsky's opinion, radio emission of main sequence RS is caused by
processes in their central regions that lead to the continuous emergence of a large number of high-energy 
charged particles ($e^-/p$) capable of emitting by the synchrotron mechanism in magnetic fields amplified near 
the RS centre. The power of this process, as Shklovsky believed, is due to a large concentration of population~II 
stars to mid-distance galactic nuclei. In the nuclei of nearly spiral galaxies the power of the process is much lower. 
In nearby galaxies a more efficient mechanism can be the formation of relativistic particles at supernova (SN) explosions. 
Possibly, in Shklovsky's opinion, this could also be applicable to 
Sgr A
. (Let us not forget that in the early 1960s 
there was yet no mention of massive accreting black holes (BH) in galactic nuclei.)

Analyzing the properties of RS on the giant sequence, Shklovsky emphasizes that 
they feature a rapid decrease of $L_{\rm r}$ with the increasing linear sizes and a non-coincidence of the 
maximum-emission regions in the optical and radio-frequency ranges. From this, he concludes that  
on this sequence RS (of the type of 
Cyg~A 
or 
Cen~A
) eject high velocity 
clouds  of gas, relativistic particles and magnetic fields with plasma
from their optical centers, i.e. from galaxies. These clouds will in 
time expand, which will lead to a decrease of their radio luminosity according to the law 
$L_{\nu}\sim R_{\nu}^{-\beta}$, where $\beta=2(2\alpha_{\rm r}+1)$, and $\alpha_{\rm r}$ is the spectral index \citep{sh4}. 
Average $\bar{\alpha}=0.7$ yields the
observed dependence 
$L_{\nu}\sim R_{\nu}^{-4.8}$. (In English-language literature, a similar model was termed the van der Laan 
model \citep{zaa}, though it was Shklovsky who first proposed it.) Herefrom, Shklovsky concluded that the 
 giant sequence may be an evolutionary one. 
 Analyzing the further fate of plasmons 
ejected from galactic nuclei, ISS believes that some of them, being decelerated in the galactic gravitational 
field, may fall back on the galaxy if their velocity is smaller than
$v_{\rm escape}$. By his estimate, the characteristic times of incidence shall be $\sim$$10^8$~years, and over that time 
plasmons can be strongly expanded diffusionally.

In the end of his paper, Shklovsky proposes a new hypothesis on the nature of radio galaxies, the data 
on the appearance of the profiles of the forbidden line [OII] 3727A in the spectrum of the central 
region of the closest radio galaxy Virgo A (M87, NGC 4486). From the fact of the presence of a large 
amount of gas both in the nucleus and in the radio ejection from this RG, Shklovsky arrives at a 
conclusion that the replenishment requires it to be delivered to the nucleus at a rate of several dozens of 
$M_{\odot}/\rm{year}$. In his opinion, such an amount of gas can be ``captured''
from intergalactic medium, when the galaxy collides during its motion with giant gas nebulae. Herewith, 
an energy of 
$\sim$$10^{43}$ erg/s 
both in kinetic and magnetic forms will be delivered to the galactic centre. If the clouds are magnetized, 
conditions are created for acceleration of charged particles to relativistic energies due to a Fermi mechanism, 
when particles intersect the region of shock waves several times by reflecting from a 
``magnetic wall''. Thus, 
Shklovsky tries to explain why some massive galaxies may become strong RS from time to time. 

\section{Extended radio sources}     
\begin{figure}      
\includegraphics[width=0.9\textwidth,angle=0]{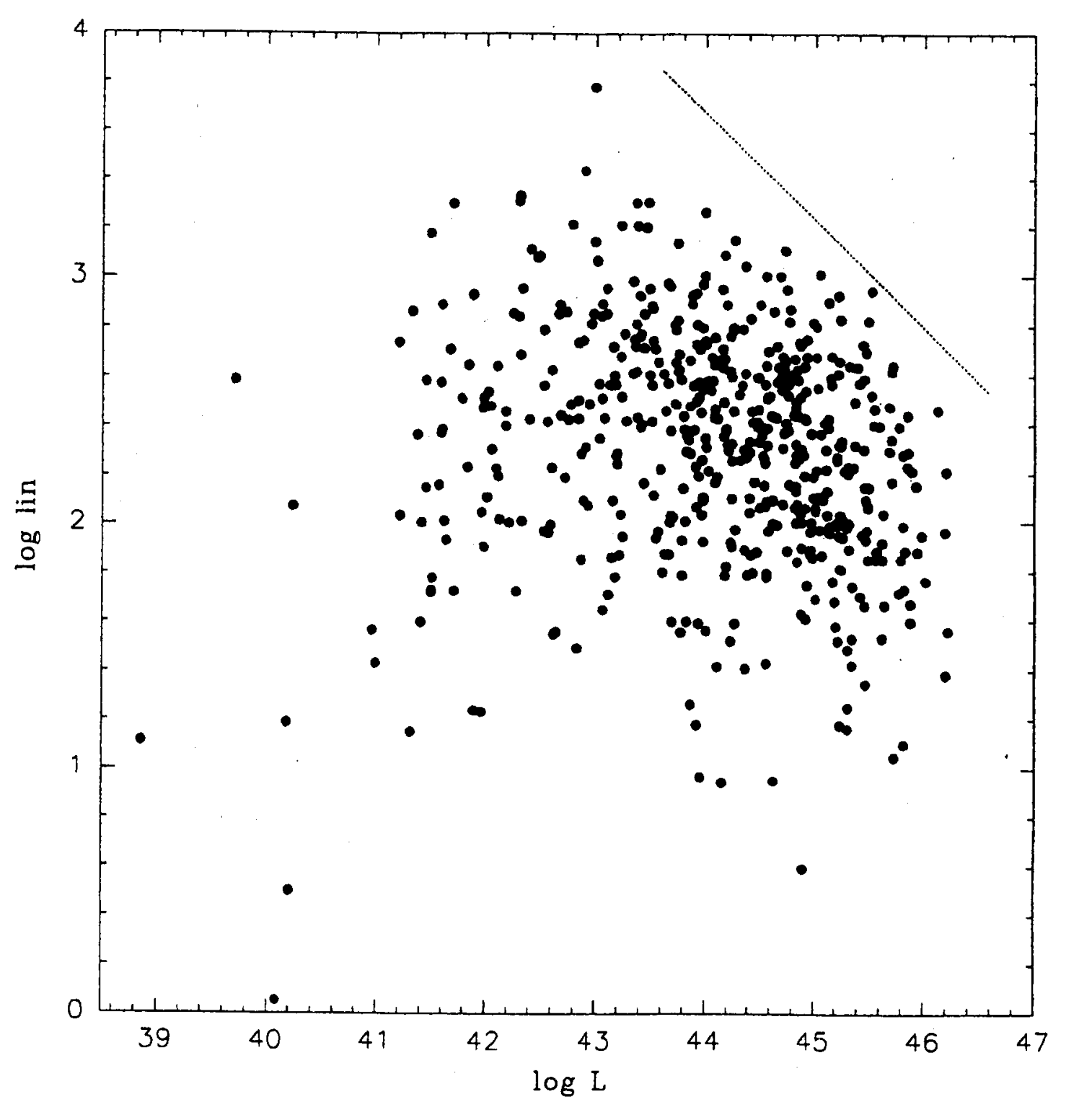}
\caption{Linear size (lin) in the dependence on $L_{\rm r}$ (10~MHz -- 10~GHz) for 267 RG and 273
 QSS with FRII radio morphology \citep{nil}.}
\label{f3}
\end{figure}

\begin{figure}      
\includegraphics[width=0.9\textwidth,angle=0]{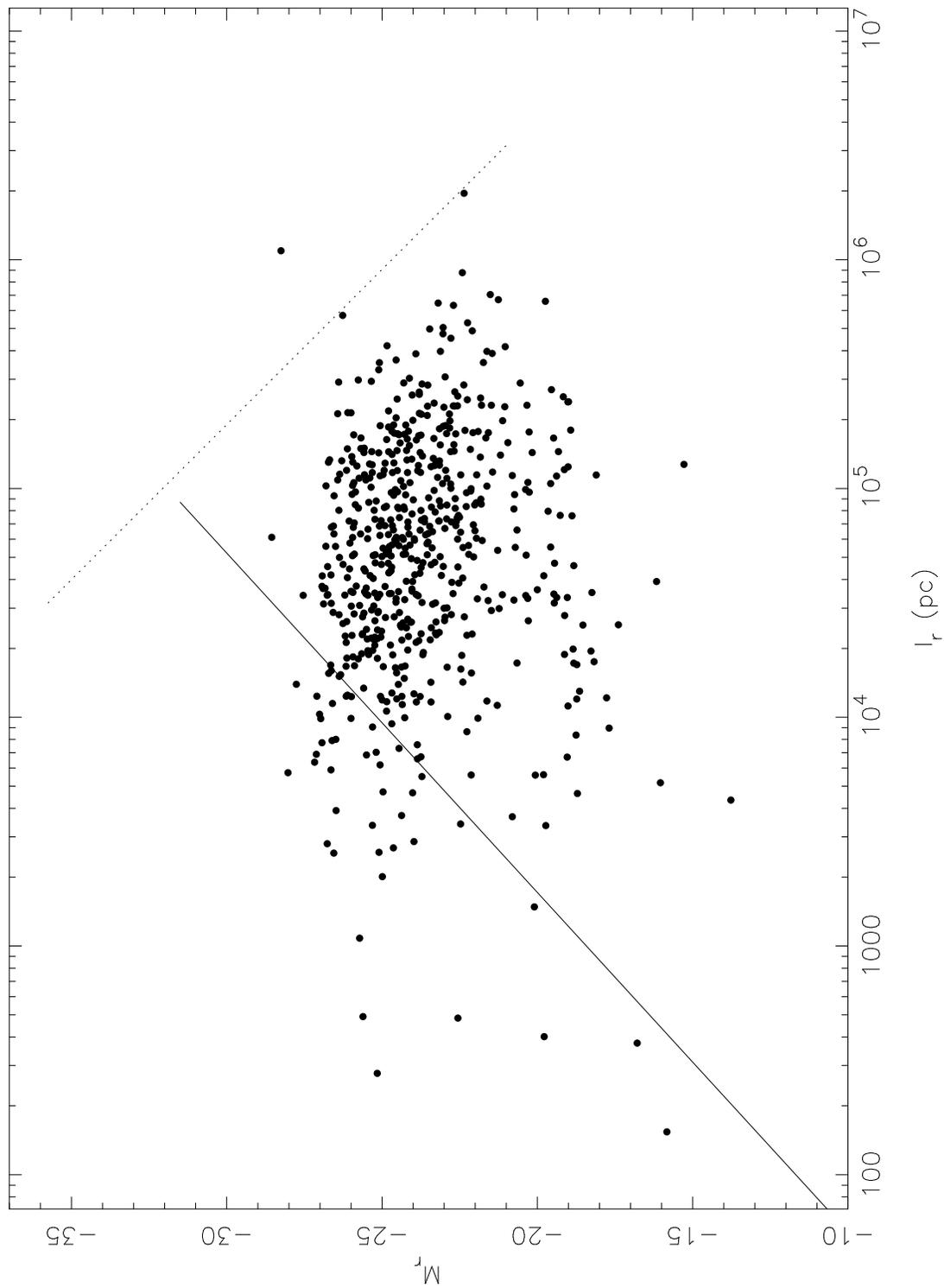}
\caption{
Absolute radio magnitudes at 178~MHz ($M_{\rm r}$) as a function of linear sizes ($l_{\rm r}$) for a 
selection of 500 RG from the 3C and 4C catalogues. The solid line shows 
the main sequence; the dashed line, the giant sequence 
\citep{kar}.}
\label{f4}
\end{figure}

In subsequent years, many authors were turning to constructing the dependence $L_{\rm r} - R_{\rm r}$ on a significantly 
larger observational material, including not only RG but also radio quasars (discovered in 1963). For instance, 
in \cite{nil} this dependence was constructed based on the data for 
$L_{\rm r}$ (10~MHz -- 10~GHz) and $R_{\rm r}$ already for 267 RG and 273 quasars (QSS) that belong by their morphology to FRII 
type (using the classification of \cite{fan}. In Fig.~\ref{f3}, it can be seen that only weak traces 
were left from the sequences about which ISS spoke. In \cite{kar}, a similar dependence was 
also constructed, based on $\sim$500~RS of various types from the 3C and 4C catalogues (Fig.~\ref{f4}). 
It is seen that there are traces of the main sequence, and the giant sequence 
is completely blurred. 
In \cite{sin},  it was noted that the sequences were blurred because the value of 
$R_{\rm r}$ is itself a function of $L_{\rm r}$ and RS red shifts. The author adduces the shape of these dependences 
obtained by 800 FRII-type RS: 
\begin{eqnarray}
 R_{\rm r} = R_{\circ}\left(\frac{P}{P_{\circ}}\right)^{\beta}(1+z)^n  \; ,
\label{one} 
\end{eqnarray}
where $P_{\circ}=10^{26.5}$~Wt/Hz, and $R_{\circ}$ are different for RG and QSS (what is more, for QSS 
$\bar{R_{\circ}}$ is smaller than for RG),
\begin{eqnarray}
\begin{tabular}{cr}
$
\qquad\qquad  \beta =
\begin{cases}
 -0.23 \quad {\rm for} \quad {\rm QSS}\\
\quad  0.35 \quad {\rm for} \quad {\rm RG}
 \end{cases} 
$ & and \quad 
$
 n =
\begin{cases}
 -0.1 \quad {\rm for} \quad {\rm QSS}\\
\quad -3 \quad {\rm for} \quad {\rm RG}.
 \end{cases} 
$\\
\end{tabular}
\end{eqnarray}
It should also be pointed out that for extended RS the value of $R_{\rm r}$ 
shall depend on the angle of ejection of radio components to the line of sight, which is poorly constrained 
based only on the ratios of radio fluxes of particular components or the distances of their centres from the 
host galaxies. To circumvent this indeterminacy, \cite{kom} instead of  $R_{\rm r}$ made use of $D_{\rm r}$, 
the diameter of the extended radio component, which depends on the orientation much weaker but, in contrast to $R_{\rm r}$, 
depends stronger on  frequency: $D_{\rm r}\sim \nu^{-0.3}$. When sampling the parameters of 
particular extended radio components, a lower bound of their sizes was introduced ($>$15~kpc) 
so that more compact ``hot'' radio spots occurring in some extended components did 
not get into the sample. Using the data on 55 individual radio components belonging to $\sim$45 different 
RS on $z<0.1$, the following dependence was obtained: 
\begin{eqnarray}
\ln \Sigma_{\nu} = -\beta \ln D_{\nu}(рс) + \ln A,
\label{two} 
\end{eqnarray}
where
\begin{eqnarray}
 \Sigma_{1,2}(\nu)=\frac{F_{\nu}}{\displaystyle\frac{\pi}{4}\left( \theta_{\nu}^{\parallel}+\theta_{\nu}^{\perp}\right)} \quad \left( \frac{{\rm Jy}}{\rm {arcsec}^2} \right)
\label{three} 
\end{eqnarray}
is the surface brightness,
$\beta=2.5\pm 0.28$, $A=9.81\pm 1.14$ and the correlation coefficient $k=0.82$ (Fig. 5).

If individual extended RS components are considered as expanding plasmons consisting of magnetic fields and 
energetic particles that lose energy on synchrotron radiation, then, as pointed out by ISS in  \cite{sh5},  
a certain relation should exist between surface radio luminosity and size of these formations ($D_{\nu}$):
\begin{eqnarray}
\Sigma_{\nu} = \frac{F_{\nu}}{D_{\nu}^2}\sim D_{\nu}^{-4(\alpha_{\rm r}+1)},
\label{four} 
\end{eqnarray}
which at $\bar{\alpha_{\rm r}}=0.75$ has the form 
\begin{eqnarray}
\Sigma_{\nu}  \sim D_{\nu}^{-7},
\label{five}   
\end{eqnarray}
i.e. $\beta=7$.
The theoretical value of $\beta$ 
does not correspond to the value obtained by \cite{kom}. 
\cite{cax} estimated these values  for remnants of SN explosions: 
\begin{eqnarray}
\bar{\beta}=
\begin{cases}
3.4\pm 0.5 \; \;  \text{for shell-like SN remnants}\\
2.35\pm 0.2  \; \text{for plerion SN remnants with embedded young pulsars.}\\
 \end{cases} 
\label{six}
\end{eqnarray}
It can be thought that, as in plerions, where energetic particles continue to be delivered to nebulae from 
young pulsars, in the same way extended radio components in RS may continue to be supplied with relativistic 
particles from the nucleus via ejections. That is, plasmons cannot be considered as merely passive expanding 
formations after their ejection from the nucleus. Besides, it is not to be ruled out that the distribution of 
relativistic particles by velocities is not isotropic, which leads to an anisotropy of emission from  radio 
components. In this case, the dependence of the value of magnetic field on $D$ can not be simply $H\sim D^{-2}$, 
and for its agreement with the observed magnitudes of $\bar{\beta}$ it should be admitted that $H\sim D^{-0.6}$.

Speaking of the work of 1962 by ISS concerning the nature of radio galaxies, one should not forget that 
radio quasars had not been discovered by that time. If the parameters of QSS are placed onto the dependence 
of $L_{\nu}\sim R$,  they will occupy the upper left margin of the giant sequence
on the plot (Fig.~\ref{f1}), 
as, on average, they are of smaller radio size than RG but are more luminous. Herewith, the same way as RG on 
the left margin of the plot, QSS by their morphology belong to FRII  and, unlike FRI-type RG, avoid the regions 
occupied by rich clusters of galaxies, at least at 
 $z<0.5$. FRII-type RS extended radio components  during their fly-off  have to overcome the resistance of interstellar 
 medium in massive coronas of elliptic galaxies, to which these RS are, as a rule, related. Thus, the fly-off 
 of these components and their expansion is decelerated not by the gravitational counteraction of the 
 host galaxy but rather by the resistance of the medium. This is manifested even more in the case of FRI-type RS that 
 are often found in clusters of galaxies. If a cluster has not yet been fully virialized and continues to compress 
 with time, FRI-type extended components may compress together with the gas of the clusters, decreasing their 
 size with time and reducing their luminosity as relativistic particles lose their energy due to synchrotron 
radiation. Therefore, FRI-type RS may evolve  down along the main sequence
in the same way as FRII-type ones, 
down the giant sequence. 

\section{Radio sources associated with active galactic nuclei}     
If one does not restrict oneself to the consideration of only the nature of strong radio sources (RG and QSS) and 
turns to the possible causes of the observed diversity of types of galaxies with active nuclei (active galactic 
nuclei, AGN) (see, e.g., \cite{n8}), 
then most authors relate this phenomenon to the diversity of the ways for formation and evolution of host galaxies 
and their central regions (bulges, near-nuclear stellar clusters, accretion disks, massive black holes). Within 
the framework of the hypothesis that the activity of nuclei is determined by the existence of  massive accreting 
BH in them, some authors, as Shklovsky, constructed for AGN analogues of HR diagrams 
(e.g., \cite{n9,n10}).
Only instead of $L_{\rm r}$ they plotted the accretion rate on the central BH ($\dot{m}$), 
and instead of size, the ratio of observed luminosity to Eddington luminosity 
($\displaystyle L_{\rm{edd}}=1.38\cdot 10^{38}\frac{M_{{\rm BH}}}{M_{\odot}}\frac{\rm{erg}}{\rm{s}}$). Such 
histograms have clear-cut regions occupied by AGN with different  luminosity ratios in the radio and optical bands 
(Fig.~\ref{f6}). This is due to the fact that, as it follows from the simplified theory of accretion of nonmagnetized 
matter on a compact body, there is some critical value 
$\dot{m}_{\rm{crit}}=0.01 \dot{m}_{\rm{edd}}$, where  $\displaystyle \dot{m}_{\rm{edd}}=\frac{L_{\rm{edd}}}{c^2}$. 

At $\dot{m}>\dot{m}_{\rm{crit}}$, an accretion disk becomes optically thick and emits a quasi-thermal spectrum with 
effective temperature  $T\sim (M_{\rm{BH}})^{-1/4}$. At $\dot{m}<\dot{m}_{\rm{crit}}$,
the character of accretion flow on the BH changes to advective flow, and the disk becomes optically thin. Herewith, 
near the BH, conditions arise for acceleration of charged particles to relativistic velocities -- a strong radio 
source emerges. That is, one and the same object  in different epochs can, depending on BH accretion rate, be 
observed as a radio source or an optical source. Herewith, mass 
 $M_{{\rm BH}}$ 
 is merely a scaling parameter and, therefore, similar estimates are valid for objects with black holes of 
 galactic and stellar masses, to which the so called microquasars 
($\mu$QSO)
belong. Studies of the  light curves change in time in $\mu$QSO in  (soft and hard) X-ray and radio-frequencies showed fast 
($\sim$$10^3$~sec) 
changes of ``states'' in them. The essence of these changes is that a clear-cut correlation 
is observed between 
the intensities of radio fluxes and the ratios of the fluxes in soft and hard X-ray range 
(see, e.g., \cite{n11}): 
when a strong emission in soft X-ray range  ($<$$0.5$~keV) is observed, radio emission is weak 
(the mode of common accretion), and when emission in hard X-ray range increases ($>$$5$~keV), strong radio emission 
emerges (the mode of advection). This behaviour is similar to the one observed in emission from active 
galactic nuclei, if we take into account that accretion disks around massive BH have a maximum of emission 
in a different spectral range  compared to disks around stellar-mass BH. Naturally, 
the energetic, temporal and spatial parameters in galactic nuclei are larger than the corresponding parameters in 
stellar active systems corresponding to the masses of their black holes, i.e., they are $10^{7-8}$ times 
greater. An indirect confirmation of such a situation can be the observational fact that both AGN and 
$\mu$QSO get to one plane, the so called fundamental plane (FP), when a correlation of not two but three 
parameters is sought for. For instance, according to the data by \cite{n12}
and \cite{n13}, 
one of the projections of this FP has the form: 
\begin{eqnarray}
\log L_{\rm{rad}} \left(\frac{\rm{erg}}{\rm{s}} \right) = 0.6\;  \log L_{\rm{XR}} + 0.78\; \log M_{\rm{BH}} .\quad  (\text{Fig. \ref{f7}})
\label{seven}
\end{eqnarray}

We would also like to note that, ISS repeatedly returned in his works to the 
nature of the activity in galactic nuclei. In those works, he put forward quite a few ideas and insights. 
Herewith, he often drew on the analogy with active processes in stars. For instance, in \cite{sh6} 
he was proving that synchrotron radiation was responsible for the optical radiation of the ejection from 
the nucleus of a massive E galaxy in the centre of a galactic cluster in Virgo A (NCC 4486). 
(Also, in 1953 Shklovsky proved that optical emission from the Crab Nebula is a continuation of 
its synchrotron radio emission.) In the same work he came up with an idea that globular clusters, 
whose spatial density in central massive E galaxies is much higher than in disk ones, can be responsible 
for the delivery of gas to galactic nuclei. It is for this reason, in the opinion of ISS, that namely 
spheroidal massive galaxies are strong RS. By Shklovsky's estimate, globular clusters may collide with 
the NGC 4486 nucleus every 
$10^6$~years, and between themselves, every 
$10^8$~years. As exemplified by observational data on the activity in the nucleus of the central galaxy 
of a cluster in Perceus NGC 1275 (Per~A), he assumed the presence of hidden quasars there  
\citep{n15,n16}. They do not manifest themselves in the optical range (most likely, 
due to the absorption of the optical-range emission by dust) but may exhibit themselves in the 
X-ray or millimetre ranges. Shklovsky, well back 50 years ago, put forward a hypothesis that is 
not generally recognized even at present that the phenomenon of activity in galactic nuclei is of 
recurrent character and that a quasar represents an active phase in the evolution of the nucleus of 
a massive E galaxy, and the nucleus of a Seyfert galaxy, an active phase in the nucleus of a disk galaxy. 

ISS in his works paid special attention to the analysis of observational data on the central region of the 
closest RG NGC 4486 (Virgo A), situated in the cluster centre in Virgo A. He frequently emphasized that for 
the problem of nuclei activity this galaxy is the same reference standard as the Crab Nebula is for the 
problem of SN explosions. As information about the activity in NGC 4486 accumulated, Shklovsky's views on 
the ways of gas delivery to its centre changed. Thus, while in \cite{sh6} 
he draws attention to the possible role of globular clusters in this process, in later works 
\citep{n17,n18}, when discussing the nature of a gaseous counter-jet and a fan jet from the nucleus of 
M87, 
he already notes the possible role of numerous planetary nebulae (PN) with small angular momenta in the delivery of gas to its 
 centre. In his opinion, the gas discharged by PN in the process of stellar evolution may produce a 
rotating massive magnetized plasma formation (of the sort of a magnetoid) in the centre of the galaxy, 
which, losing its stability, may explode. Proceeding from these model views, ISS considered the activity 
phenomenon in the nuclei as a consequence of stellar evolution processes in the central regions of massive 
galaxies. And the fact that only 10--20\% of external galaxies are strong RS was explained by him as an 
argument in favour of the recurrence of this phenomenon. 

Shklovsky perceived a profound analogy between processes leading to the ejection of plasmons from active 
nuclei with a large spread of velocities (up to relativistic ones), which are decelerated in the external
 medium of host galaxies, and processes occurring in remnants of SN explosions, which also expand in the 
 external medium (see, e.g., 
 \cite{n19,n20}). By his estimates, similar to the ones he made for SN remnants, nuclear plasmons have 
 masses from 1 up to  $10^3M_{\odot}$, the on/off ratio of their ejections determined by the accumulation 
 of gas in the nuclei is about several thousand years and the efficiency of the transfer of their kinetic 
 energy ($\sim$$10^{54}$~erg) to the energy of relativistic particles does not exceed 10\%.

In some of his works (e.g., \cite{n21,n22}) ISS pays attention to a distortion of the real pattern of 
relativistic radio ejections due to special-relativity effects. In this case
\begin{eqnarray}
\displaystyle\frac{v_{\rm{obs}}}{c}=\displaystyle\frac{\left(\displaystyle\frac{v_{\circ}}{c} \right)\sin\Theta}{1-\left( \displaystyle\frac{v_{\circ}}{c}\right) \cos\Theta} \quad \rm{and} \quad  \textit{L}_{\rm{obs}}=\textit{L}_{\circ}\delta^{\textit n}, \quad \textit{n}=2\!-\!3;
\label{eight}
\end{eqnarray}
where the Doppler factor
\begin{eqnarray}
\delta=\left[ \gamma\left( 1-\frac{v_{\circ}}{c}\cos\Theta\right)\right]^{-1} 
\label{nine}
\end{eqnarray}
and the Lorentz factor
\begin{eqnarray}
 \gamma=\left [1-\left( \frac{v_{\circ}}{c}\right)^2\right]^{-1/2}.
\label{ten}
\end{eqnarray}
These effects, Shklovsky maintained, can explain the one-sided character of radio ejections if they are 
directed under small angles $\Theta\lesssim 10^{\circ}$ to the line of sight. (As it now becomes clear, 
in objects of the type of BL~Lac this angle can be even smaller, which  explains their fast variability 
and relatively high radio luminosity.) It is true, though, ISS noted, that this explanation is not suitable 
for Virgo A RG, because the velocity of its radio ejection is 
$\sim 0.1c$ \citep{n23} 
and is, therefore, one-sided in reality, though it would seem to contradict the counter-jet observed in the 
nucleus of NGC~4486. Even today this problem still remains not fully understood, though some works proposed 
a flip-flop model, when ejections fly off from the nucleus to opposite sides but in different phases of 
activity (e.g., 
 \cite{ens,zhu}). In this case, in a given epoch an ejection will be observed to be to one side, and in 
 the next epoch, to the other. But extended long-lived radio components can be observed to be on both 
 sides of the host galaxy. 
 
In the subsequent years of his life, ISS became interested in the problem of the nature of nearby optical 
quasars: are they really young objects or are they old but, for some reason, with the delayed  phase of 
activity (which for quasars does not exceed $10^7$ years)? For radio quasars, an answer to this question 
could be given by observations of old extended radio components in them, which would testify to the 
recurrent character of radio ejections from them, but for optical quasars some other explanation should 
be sought for. Shklovsky has suggested that nearby quasars are old objects in which only in the present 
epoch conditions emerged for the formation of a near-nuclear disk and an intensive accretion of gas to 
the nucleus began. Why it could not have happened earlier, remained not clear. 

\section{Conclusion}

It is only to be regretted that Iosif S. Shklovsky had not lived at least a dozen of years more -- how 
many of his brilliant ideas and hypotheses we would have come to know about. And, most important of all, 
he would have seen with his own eyes all-wave astrophysics he dreamed about, for the formation of which he 
did so much. 


\newpage

\newpage

\newpage

\newpage

\newpage
\begin{figure}      
\includegraphics[width=0.9\textwidth,angle=-90]{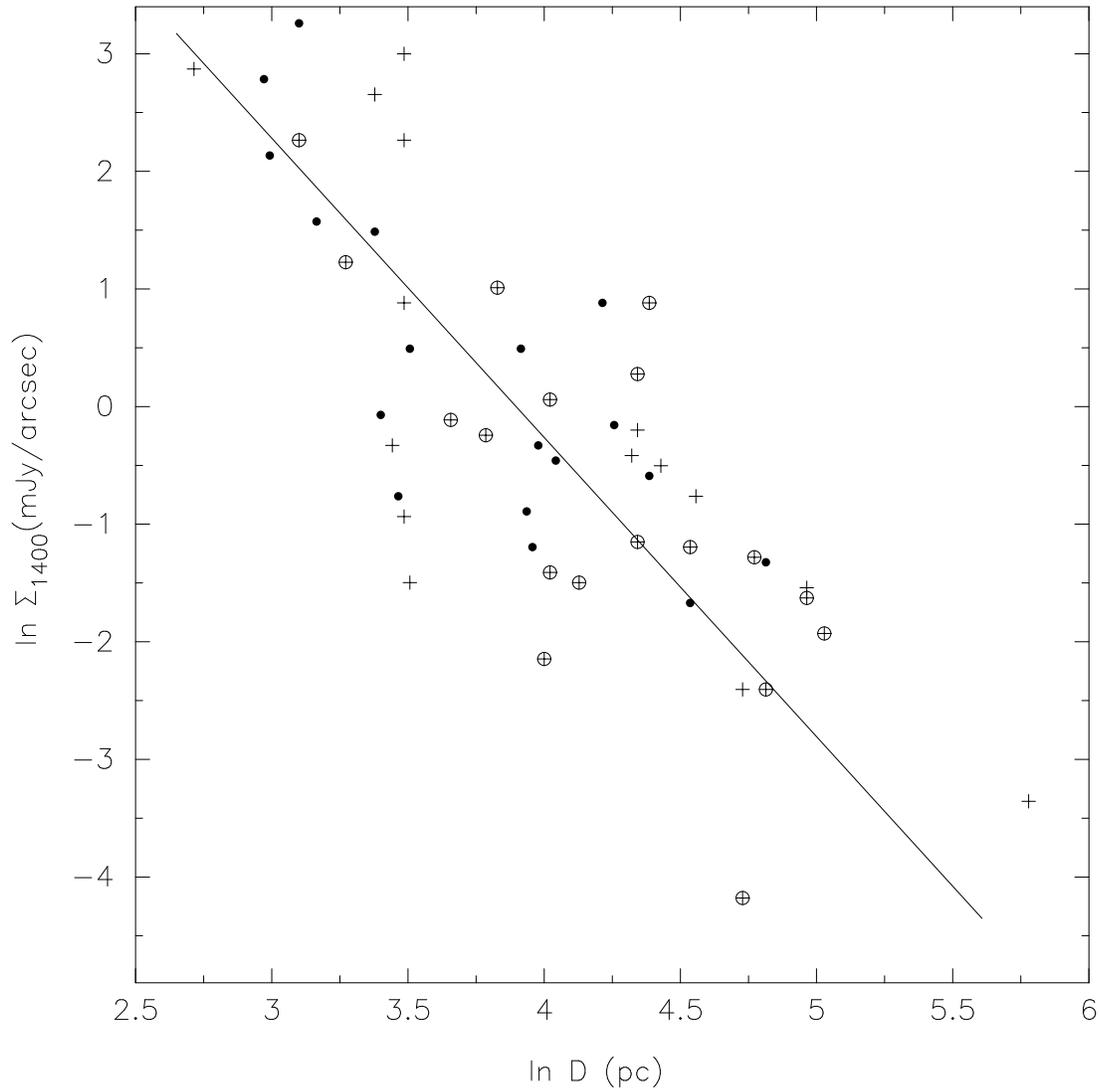}
\caption{Dependence of $\sum_{\nu}$ on $D_{\nu}$ at 1400~MHz for 55 extended radio components of close radio galaxies  ($z<1$).
Correlation coefficient, $\sim$0.82; slope of the dependence, $-2.5\pm0.5$. Components: $\bullet$, small; $\oplus$, large; $+$, equal components (or one resolved  component) \citep{kom}.}
\label{f5}
\end{figure}

\newpage
\begin{figure}     
\includegraphics[width=1.0\textwidth,angle=0]{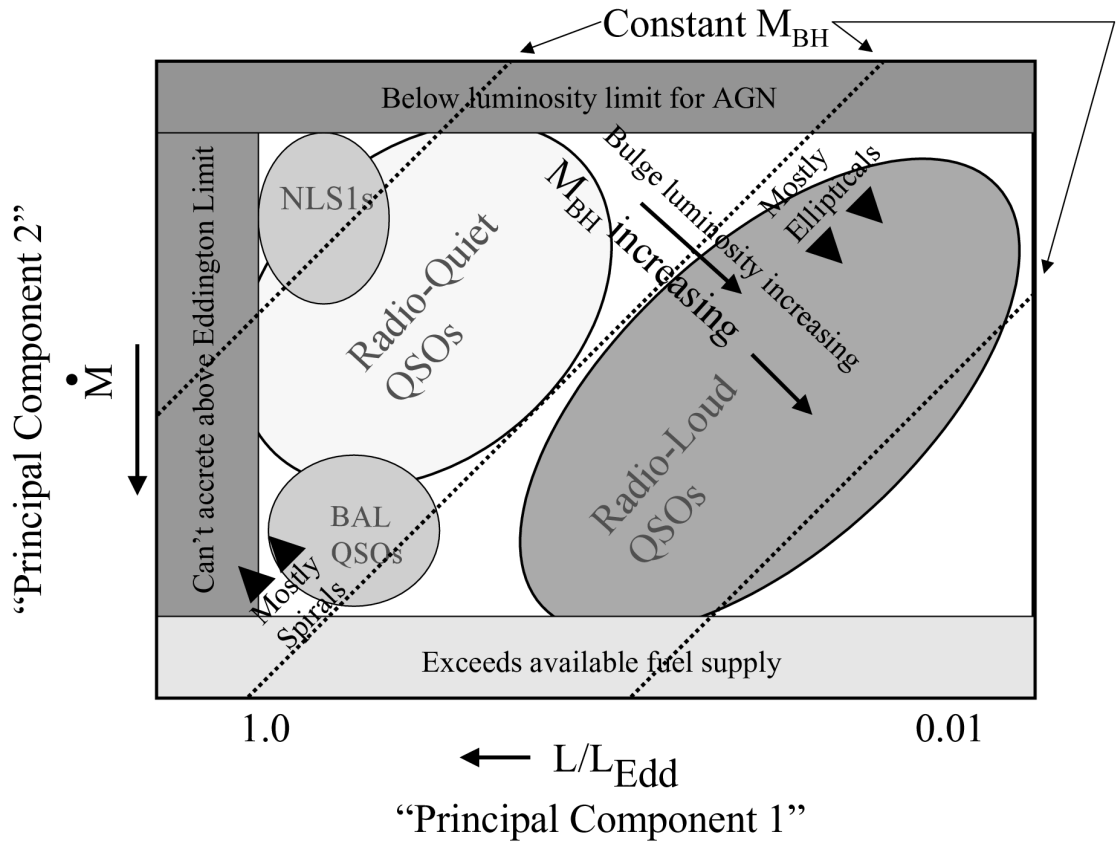}
\caption{Classification of AGN on the plane of PC1 vs PC2 (an analogue of the HR diagram for AGN). PC1 
reflects the halfwidths of $H_{\beta}$ lines in AGN spectra, and, therefore, $M_{\rm {BH}}$, i.e., 
luminocity; PC2 reflects the relation of the FeII equivalent widths 
4570\,\AA \; and $H_{\beta}$, i.e, in a sense, accretion disk temperature \citep{n9}.} 
\label{f6}
\end{figure}

\newpage
\begin{figure}    
\includegraphics[width=0.75\textwidth,angle=-90]{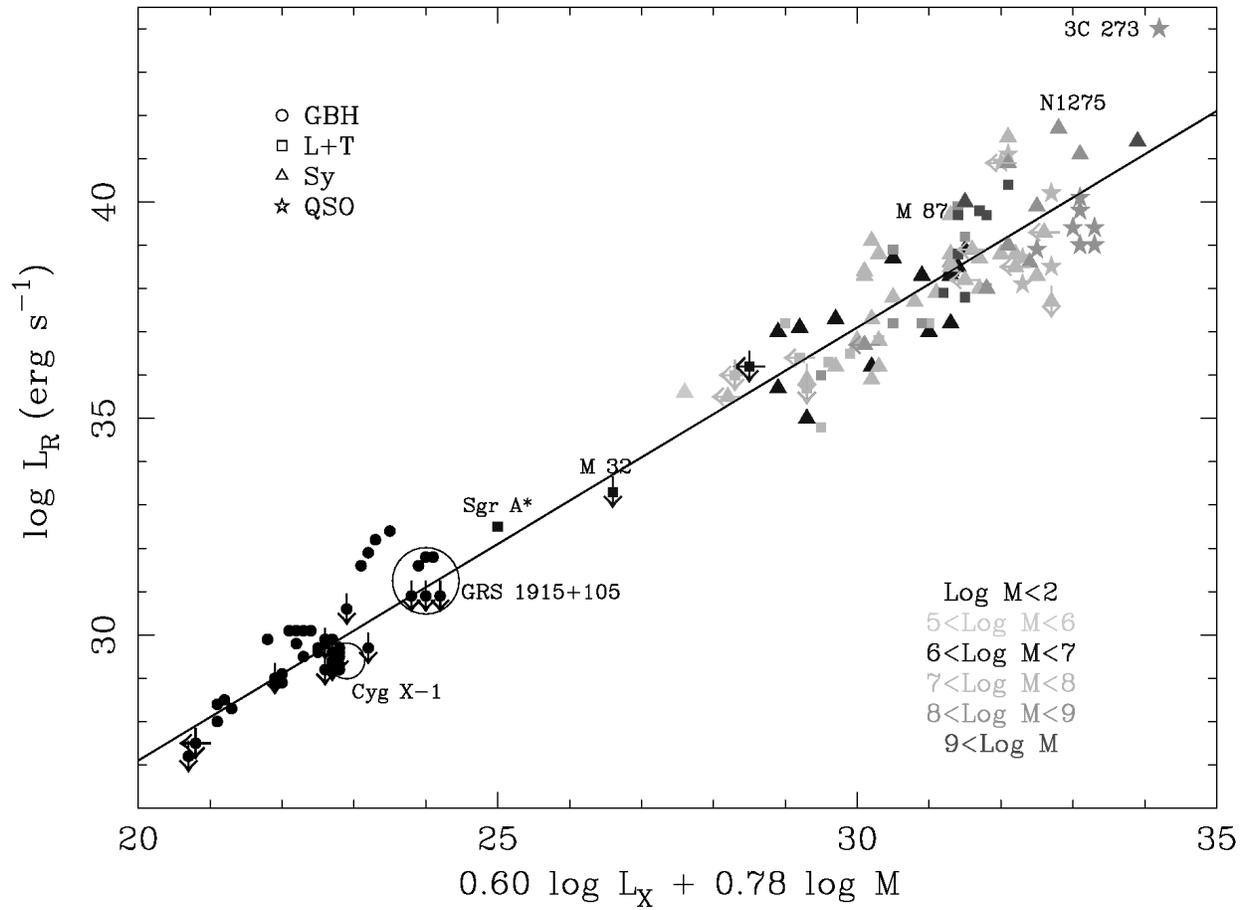}
\caption{One of the projections of the fundamental plane \citep{n12}, on which both active stellar systems ($\mu$QSO) and active galactic nuclei fall.}
\label{f7}
\end{figure}

\end{document}